\documentclass[showpacs,preprintnumbers,superscriptaddress,amsmath,amssymb,nofootinbib, floatfix]{revtex4}
\usepackage{epsfig,graphics,color,graphicx,amsmath}
\usepackage[section]{placeins}

\usepackage[usenames,dvipsnames]{xcolor}
\usepackage[normalem]{ulem}

\newcommand{\cd}{\makebox[0.08cm]{$\cdot$}}

\newcommand{\sla}{\not\!}
\begin{document}
\title{Transition electromagnetic form factor and  current conservation  in the Bethe-Salpeter approach}

\author{J.~Carbonell}
\affiliation {Institut de Physique Nucl\'eaire,
Universit\'e Paris-Sud, IN2P3-CNRS, 91406 Orsay Cedex, France}
\author{V.A. Karmanov}
\affiliation {Lebedev Physical Institute, Leninsky Prospekt 53, 119991
Moscow, Russia}

\bibliographystyle{unsrt}

\begin{abstract}
The transition form factor for electrodisintegration of a two-body bound system is calculated in the Bethe-Salpeter framework. 
For the initial (bound) and the final (scattering) states, we use our solutions of the Bethe-Salpeter equation in Minkowski space which were first obtained recently.
The gauge invariance, which  manifests itself in the conservation of the transition electromagnetic current $J\cdot q=0$, is studied numerically. 
It results from a cancellation between  the plane wave  and the final state interaction contributions. 
This cancellation takes place
only if the initial bound state BS amplitude, the final scattering state and the operator of electromagnetic current are strictly consistent with each other, that is if
they are found in the same dynamical framework.  A reliable result for the transition form factor can be obtained in this case only.
\end{abstract}

\maketitle
%

\section{Introduction} \label{intro}

Computing  the electromagnetic (EM) form factors in the Bethe-Salpeter (BS) approach \cite{bs} requires the solutions of the BS equation in Minkowski space. 

The main reason is that the Wick rotation \cite{WICK_54}, which allows to go from Minkowski to Euclidean space in the BS equation, cannot be performed in the integral expression 
of the EM form factor (see e.g. \cite{ck-trento}). 
However, in contrast to the Euclidean case, finding the Minkowski space solution is complicated by the many singularities in the integrand of the BS equation
and in the amplitude  itself. In the recent years, and using different independent methods, these difficulties have been overcome and an important progress was achieved.

In one of these methods \cite{burov},  the kernel of the BS equation is approximately represented in a separable form. This allows to considerably advance analytically and therefore simplifies finding the solution. 

In the method developed in refs. \cite{KusPRD95,KusPRD97}  the  BS amplitude is represented as an integral 
over a weight function  $g$ -- the so called Nakanishi transform \cite{nak63} --  which satisfies a nonsingular equation.  
A modification of this method, based on  the light-front projection of the BS amplitude,  was developed   in \cite{bs1,bs2,Sauli_JPG08,FrePRD14,GianniFBS2014} and used to find the bound state  Minkowski BS amplitude. 
The elastic EM form factor was also calculated in \cite{ckm_ejpa}.  
An equation for the Nakanishi function for the scattering states was derived  \cite{fsv-2012}.

\begin{figure}[hbtp] 
\begin{minipage}{17.cm}
\begin{center}
\includegraphics[width=7.cm]{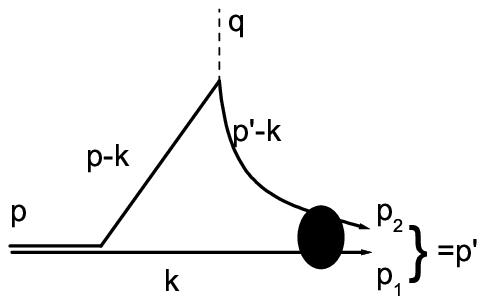} \hspace{0.8cm}
\includegraphics[width=7.cm]{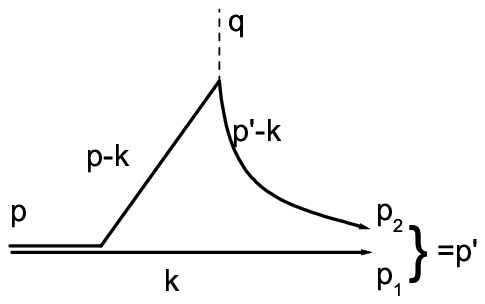}
\caption{Left panel: Feynman graph for contribution of FSI to the transition EM form factor.
Right panel: PW contribution to  the transition EM form factor.}\label{triangle}
\end{center}
\end{minipage}
\end{figure}

Another method  \cite{LC2013-2,CK_PLB_PRD}  is based on the direct solution  in Minkowski space of the BS equation after an appropriate treatment of singularities. 
The scattering problem was there solved and the off-mass shell scattering amplitude  first computed. 
Once reduced to the mass shell, the latter reproduces the phase shifts. 
Taken off-mass shell, it allows to calculate the electrodisintegration of the bound system, i.e.  the form factor of the transition bound  to scattering state.

An important contribution to this form factor, incorporating the final state interaction (FSI), is given by the Feynman graph shown in Fig.  \ref{triangle} (left panel). The right and left vertices in this graph are just 
the Minkowski BS amplitudes for the bound (left) and scattering (right) states. 
If both vertices correspond to a bound state (case of the elastic form factor), the Nakanishi transform allows to calculate the 
4D Feynman integral corresponding to Fig. \ref{triangle} (left panel),  with an integrand containing three singular propagators, analytically \cite{ckm_ejpa}. Then the non-singular integral with the weight functions $g$ is safely calculated numerically. 
For the scattering state, though the Nakanishi transform also exists \cite{fsv-2012}, the corresponding weight function $g$ at positive energies is not yet computed. Note, however, that very recently $g$ was found in the zero-energy limit that allowed to calculate the scattering length \cite{fsv_a0}.
Without using the Nakanishi representation the scattering state vertex can be  obtained only numerically \cite{CK_PLB_PRD} and therefore the singular 4D Feynman integral corresponding to Fig. \ref{triangle} (left panel) must  be  computed numerically as well.
This calculation,   providing the transition electromagnetic current and the form factor,  requires however some care to take properly into account  the pole singularities of the propagators. 

The aim of this paper is to give the detail of  the first results presented in  \cite{transit,LC2014} and analyze the conservation of the calculated electromagnetic current in the inelastic transition. 
We will see that this current  is indeed conserved, as it should be from general principles \cite{Adam_V_Orden_Gross}. 
However this conservation is due to a rather delicate cancellation between the plane wave (PW) contribution (right panel of Fig. \ref{triangle}) and the  final state interaction (left panel of Fig. \ref{triangle})
which requires a strict consistency between, on one hand, the bound and scattering state solutions and, on the other hand, 
the electromagnetic current operator. It thus provides a strong test for all these quantities simultaneously.

The need for an internal consistency between  states, currents and dynamical equation to ensure the gauge invariance was extensively  discussed in \cite{Adam_V_Orden_Gross} 
in the framework of the BS and the Gross spectator equations. 
Our numerical results are in agreement with this general expectation. 
We will show that if this consistency and hence the gauge invariance is violated, a consequence of that is not   only   the
appearance  of a non-conserved part  in the current -- which anyway  drops out in the cross section  -- but that the current as a whole is not valid at all. 
In other words, the transition form factors extracted from the conserved part of the non-conserved current, are also deficient.

In order to illustrate our treatment of  the singularities, we will restrict to the spinless particles. 
The generalization to the fermion case is straightforward since the fermion and scalar propagators have the same singularities.

The paper is organized as follows.
In Section \ref{tff} we discuss the decomposition of the transition current  in the form factors  without assuming the current conservation. 
In Sections \ref{fsi} and \ref{pwc}, the FSI and PW contributions  in the conserved part of the current are calculated. 
Section \ref{conserv} is devoted to the  discussion of the current conservation. 
In particular, the FSI and PW contributions of the non-conserved part of the current have been calculated  and we have shown that they cancel each other. 
Some selected numerical results are presented in Section \ref{numerics}. 
Section \ref{concl} contains the concluding remarks. 
The cumbersome details of the  calculations are given in the Appendices \ref{frame}, \ref{app1} and \ref{app_PW}.

\section{Transition form factor}\label{tff}

In the case of spinless particles, and without supposing the current conservation,  the general form of the electromagnetic current involves two  form factors:
\begin{equation}\label{JF1F2}
J_{\mu}= (p_{\mu}+p'_{\mu}) F_1(Q^2) + (p'_{\mu}-p_{\mu})F_2 (Q^2)
\end{equation}
The decomposition (\ref{JF1F2}), together with the scalar character of the constituents, 
implies  that the initial and final states have total zero angular momenta, i.e. that they are composed of  S-waves only.

In order to  study the current conservation, it is convenient to redefine the form factors by introducing the following linear combinations $F$ and $F'$:
\[\begin{array}{lcl}
F &=& F_1 \cr
F' &=& F_1-\frac{Q^2}{Q_c^2} F_2 
\end{array}
\qquad \Longleftrightarrow \qquad
\begin{array}{lcl}
F_1 &=& F \cr
F_2 &=& \frac{Q_c^2}{Q^2} (F-F') 
\end{array}
\]
with
\begin{eqnarray}
q_{\mu} &=&  p'_{\mu}-p_{\mu} \cr
Q^2        &=& -q^2=-(p'-p)^2   \cr
Q_c^2    &=&  {M'}^2-M^2
\end{eqnarray}
$M$ is the initial bound state mass, $M'$ is the invariant mass of the final scattering state.
In terms of them, the current (\ref{JF1F2}) can be rewritten in the form:
\begin{equation}\label{ffc}
J_{\mu}=\left[(p_{\mu}+p'_{\mu})+ (p'_{\mu}-p_{\mu})\frac{Q_c^2}{Q^2}\right]F(Q^2)- (p'_{\mu}-p_{\mu})\frac{Q_c^2}{Q^2}F'(Q^2)
\end{equation}
Since
\begin{equation}\label{CC}
q\cdot J= Q_c^2F'(Q^2)
\end{equation} 
the current conservation $q\cdot J=0$ is equivalent  to $F'(Q^2)\equiv 0$. 

Notice that in the elastic case the form factor $F'$ is absent since the term \mbox{$\sim (p'_{\mu}-p_{\mu})F'(Q^2)$} in (\ref{ffc}) is forbidden by the symmetry between initial and final states.

Notice also that the form factor $F'(Q^2)$, even if it is not zero,  does not contribute to the electrodisintegration amplitude $A$. 
Indeed, this amplitude is given by:
$$  A\sim \frac{J_{\mu}\;\bar{u}(k')\gamma^{\mu}u(k)}{Q^2}  $$
It contains the electron spinors $u(k)$ and $\bar{u}(k')$. 
Substituting here the current (\ref{ffc}) and using the Dirac equation, we see that the term containing $F'(Q^2)$ drops out since 
$$  (p'_{\mu}-p_{\mu})\bar{u}(k')\gamma^{\mu}u(k)=\bar{u}(k')(\sla{k}-\sla{k'})u(k)=0.  $$

Below we will calculate each of these form factors -- $F$ and $F'$ -- as a sum of FSI (left panel in Fig. \ref{triangle})  and PW (right panel in Fig. \ref{triangle}) contributions,  in the form:
\begin{eqnarray}
F_{inel}(Q^2) &=& F_{fsi}(Q^2)+F_{pw}(Q^2)  \cr
F'_{inel}(Q^2) &=& F'_{fsi}(Q^2)+F'_{pw}(Q^2)   \label{fffull}
\end{eqnarray}
We will check that the full current  is conserved, that is, for any $Q^2$,  the contributions to $F'_{inel}(Q^2)$ of  
the FSI  and PW cancel each other: 
\begin{equation}\label{FpQ2}
 F'_{inel}(Q^2)=F'_{fsi}(Q^2)+F'_{pw}(Q^2)=0,
\end{equation}
provided the bound and scattering states are solutions of the BS equation  with the one-boson exchange kernel.
In this case,  the EM current corresponding to the interaction of a photon with a constituent is free.
We are however  interested in a quantitative measure of the accuracy of this cancellation in a real calculation.
This is the reason for introducing in (\ref{ffc})  the non-conserved part  --  proportional to  $(p'_{\mu}-p_{\mu})F'(Q^2)$  --  and the value of the form factor $F'(Q^2)$ will give us this measure. 

We will calculate separately the FSI  and   PW  contributions  to the form factor $F'(Q^2)$  and see with what accuracy they cancel each other in  the sum  (\ref{FpQ2}).
\bigskip

\section{Final state interaction}\label{fsi}

We start with the FSI contribution. 
It  is obtained by applying the Feynman rules to the left panel graph of Fig.   \ref{triangle}
and has the form  (following the convention of \cite{IZ}):
\begin{equation}\label{ffin0}
J_{\mu,fsi}=i\int \frac{d^4k}{(2\pi)^4}\,
\frac{(p_{\mu}+p'_{\mu}-2k_{\mu})\Gamma_i \left(\frac{1}{2}p -k,p\right)\Gamma_f
\left(\frac{1}{2}p'-k,p'\right)} {(k^2-m^2+i\epsilon)[(p-k)^2-m^2+i\epsilon]
[(p'-k)^2-m^2+i\epsilon]},
\end{equation}
Here $\Gamma_i$ is the initial (bound state) vertex and $\Gamma_f$ is the final vertex (half-off-shell scattering BS amplitude). 
As mentioned, both vertex functions were found numerically by solving the S-wave BS equation  in \cite{CK_PLB_PRD}. 
More precisely the function $\Gamma_f$ is related to the scattering wave solution  $F_0$ by Eq. (\ref{GF0}) from Appendix \ref{appB4}.

The integrals of the type (\ref{ffin0}) are usually calculated by applying to the product  of propagators the Feynman parametrization and then performing the Wick rotation. 
However, besides the  product  of propagators, expression  (\ref{ffin0}) contains the initial ($\Gamma_i$)  and final ($\Gamma_f$) BS amplitudes which are known numerically.  
Therefore  the  Feynman parametrization cannot be applied and we should calculate this 4D singular integral numerically, though after some  transformations. 

\bigskip
It is convenient to carry out this calculations in the system of reference where $p'_0=p_0$ (i.e. $q_0=0$) and $\vec{p}$ and $\vec{p'}$ are collinear,
i.e. they  either are parallel or anti-parallel to each other, depending on the kinematical conditions. 
In the elastic case it coincides with the Breit frame
$\vec{p}+\vec{p'}=0$ and, one has of course $|\vec{p}|=|\vec{p'}|$,  $p'_0=p_0$. 
In the inelastic case, in the frame with $p'_0=p_0$ we have $|\vec{p}|\neq|\vec{p'}|$.  
Some useful kinematical relations valid in this reference system are given in  Appendix \ref{frame}. 
\bigskip

In this reference frame we take the zero component of the current (\ref{ffc}) and get the relation:
\begin{equation}\label{JF}
J_{0}=2p_{0}F(Q^2)
\end{equation}
That is:
\begin{equation}\label{ffin1}
F_{fsi}(Q^2)=i\int \frac{dk_0\,d^3k}{(2\pi)^4}\;
\frac{(p_0-k_0)}{p_0}\;\frac{\Gamma_i \left(\frac{1}{2}p -k,p\right)\Gamma_f
\left(\frac{1}{2}p'-k,p'\right)} {(k_0^2-\varepsilon_{\vec{k}}^2+i\epsilon)[(p_0-k_0)^2-\varepsilon_{\vec{p}-\vec{k}}^2+i\epsilon]
[(p_0-k_0)^2-\varepsilon_{\vec{p'}-\vec{k}}^2+i\epsilon]}
\end{equation}
with $\varepsilon_{\vec{q}}=\sqrt{m^2+{\vec{q}}^2}$ and   similar expressions for $\varepsilon_{\vec{p}-\vec{k}}$ and $\varepsilon_{\vec{p'}-\vec{k}}$
obtained using (\ref{pmk2}) and (\ref{ppmk2}).

\bigskip
As  detailed in the Appendix \ref{appB4}, in case of initial and final S-waves, all kinematical variables as well as the arguments
of the vertex functions $\Gamma$  appearing in (\ref{ffin1})
can be expressed in terms of  $|\vec{p}|$, $|\vec{p'}|$ and the integration variables $(k_0,z,|\vec{k}|)$ with $z=\hat{k}\cdot\hat{p}$.
To lighten the writing we will denote hereafter abusively $p=|\vec{p}|$, $p'=|\vec{p'}|$ and $k=|\vec{k}|$.  

After a trivial integration over the azimutal angle, the integration measure in (\ref{ffin1}) becomes 
\[ \frac{dk_0\,d^3k}{(2\pi)^4}= \frac{dk_0\,dz\; k^2dk}{(2\pi)^3} \]

Let us introduce the following notations, making explicit only the dependence on the integration variables:
\begin{equation}\label{f}
f(k_0,z,k)=\frac{G(k_0,z,k)} {(k_0^2-\varepsilon_{\vec{k}}^2+i\epsilon)[(p_0-k_0)^2-\varepsilon_{\vec{p}-\vec{k}}^2+i\epsilon] [(p_0-k_0)^2-\varepsilon_{\vec{p'}-\vec{k}}^2+i\epsilon]},
\end{equation}
where
\begin{equation}\label{Gk0}
G(k_0,z,k)=\frac{(p_0-k_0)}{p_0}\;\Gamma_i \left(\frac{1}{2}p -k,p\right)\Gamma_f \left(\frac{1}{2}p'-k,p'\right).
\end{equation}

Each pole singularity in (\ref{f}) is represented as a sum of its principal value and a delta-function and therefore the function $f$ takes the form:
\begin{eqnarray}\label{fk0}
f(k_0,z,k)&=&G(k_0,z,k)\left[
\mbox{PV}\frac{1}{k_0^2-\varepsilon_{\vec{k}}^2}-i\pi\delta(k_0^2-\varepsilon_{\vec{k}}^2) \right]   \nonumber\\
&&\phantom{(k_0)k}\times \left[\mbox{PV}\frac{1}{(p_0-k_0)^2-\varepsilon_{\vec{p}-\vec{k}}^2}-i\pi\delta\Bigl((p_0-k_0)^2-\varepsilon_{\vec{p}-\vec{k}}^2\Bigr) \right]   \nonumber\\
&&\phantom{(k_0)k}\times \left[\mbox{PV}\frac{1}{(p_0-k_0)^2-\varepsilon_{\vec{p'}-\vec{k}}^2}-i\pi\delta\Bigl((p_0-k_0)^2-\varepsilon_{\vec{p'}-\vec{k}}^2\Bigr) \right]   \nonumber\\
&\equiv& f_3+f_2+f_1,
\end{eqnarray}
where  $f_3$ is the contribution of the product of three principal values and no delta-functions  (one single term),  
$f_2$ is the contribution of the product of two principal values  and one delta-function (three terms) and $f_1$ is the contribution of the product of one principal value  and two delta-functions (also three terms). 
The product of three delta-functions does not contribute since their arguments cannot be zero simultaneously. 

These functions $f_{i}$ have the  following explicit form:
\begin{eqnarray}
f_3(k_0,z,k) &=&G(k_0,k)\;\mbox{PV}\frac{1}{k_0^2-\varepsilon_{\vec{k}}^2}  \;\mbox{PV}\frac{1}{(p_0-k_0)^2-\varepsilon_{\vec{p}-\vec{k}}^2} 
 \;\mbox{PV}\frac{1}{(p_0-k_0)^2-\varepsilon_{\vec{p'}-\vec{k}}^2} 
\label{f3}
\\
f_2(k_0,z,k)&=&-i\pi\delta(k_0^2-\varepsilon_{\vec{k}}^2)G(k_0,z,k) \;\mbox{PV}\frac{1}{(p_0-k_0)^2-\varepsilon_{\vec{p}-\vec{k}}^2}    \;\mbox{PV}\frac{1}{(p_0-k_0)^2-\varepsilon_{\vec{p'}-\vec{k}}^2}    \cr
&\phantom{=}&-i\pi\delta((p_0-k_0)^2-\varepsilon_{\vec{p}-\vec{k}}^2)G(k_0,z,k)
 \;\mbox{PV}\frac{1}{k_0^2-\varepsilon_{\vec{k}}^2}  \;\mbox{PV}\frac{1}{(p_0-k_0)^2-\varepsilon_{\vec{p'}-\vec{k}}^2}     \cr
&\phantom{=}&-i\pi\delta((p_0-k_0)^2-\varepsilon_{\vec{p'}-\vec{k}}^2)G(k_0,z,k) \;\mbox{PV}\frac{1}{k_0^2-\varepsilon_{\vec{k}}^2}  \;\mbox{PV}\frac{1}{(p_0-k_0)^2-\varepsilon_{\vec{p}-\vec{k}}^2}    \label{f2}     \\
f_1(k_0,z,k)&=&-\pi^2\delta(k_0^2-\varepsilon_{\vec{k}}^2)\delta((p_0-k_0)^2-\varepsilon_{\vec{p}-\vec{k}}^2)G(k_0,z,k)    \;\mbox{PV}\frac{1}{(p_0-k_0)^2-\varepsilon_{\vec{p'}-\vec{k}}^2} 
\cr
&\phantom{=}&-\pi^2\delta(k_0^2-\varepsilon_{\vec{k}}^2)\delta((p_0-k_0)^2-\varepsilon_{\vec{p'}-\vec{k}}^2)G(k_0,z,k)   \;\mbox{PV}\frac{1}{(p_0-k_0)^2-\varepsilon_{\vec{p}-\vec{k}}^2}    \cr
&\phantom{=}&-\pi^2\delta((p_0-k_0)^2-\varepsilon_{\vec{p}-\vec{k}}^2)\delta((p_0-k_0)^2-\varepsilon_{\vec{p'}-\vec{k}}^2)G(k_0,z,k) \;\mbox{PV}\frac{1}{k_0^2-\varepsilon_{\vec{k}}^2}    \label{f1}
\end{eqnarray}
The index of $f_{i}$ ($i=1,2,3$) denotes the number of the principal value products that involves.

Our task now is to calculate  the 4D integral (\ref{ffin1}), rewritten as
\[  F_{fsi}(Q^2)=\frac{i}{(2\pi)^3} \;  \int dk_0\,dz\;k^2dk   \; \left\{ f_3(k_0,z,k)+f_2(k_0,z,k)+f_1(k_0,z,k) \right\}   \]
 with $f_{i}$ given by Eq. (\ref{f3}-\ref{f1}). 
Part of this integration is calculated analytically and the remaining part, once transformed  into a non-singular integrand, numerically.

For calculating the singular principal value integrals in  $f_3$  we will use the subtraction technique. 
That is, we subtract and add to $f_3$ an appropriately chosen singular function $h_3$ which, in  variable $k_0$ has the same poles as $f_3$
and has no any other singularities: 
\[ f_3=(f_3-h_3)+h_3.\] 
In the difference $(f_3-h_3)$, the pole singularities cancel each other and the result is a smooth function,  whereas in  the additional term $h_3$ the integral over $dk_0$ is calculated analytically. 

After this calculation,  there still remains a singular expression in variable $\vec{k}$. It is however logarithmic  and can be treated by using standard numerical techniques, like variable change
or by simply increasing the number of integration points. 
The details of all these calculations are given in Appendix \ref{app1}.

In the integrals containing the  functions $f_2$ and  $f_1$,  the integration over $k_0$ is easily performed analytically by means of the delta-functions.
After that, and a trivial azimuthal integration, the result is reduced to a two- and one-dimensional numerical integrations respectively. 

The final result for the FSI contribution (\ref{ffin1}) reads:
\begin{eqnarray}\label{FQ}
F_{fsi}(Q^2)&=& \frac{i}{(2\pi)^3} \;  \int dk_0\,dz\;k^2dk\;  \; \left\{ f_3(k_0,z,k)+f_2(k_0,z,k)+f_1(k_0,z,k) \right\}
\cr
&\equiv & F_3(Q^2)+F_2(Q^2)+F_1(Q^2)
\end{eqnarray}
where $F_{i}(Q^2)$  are defined in Appendix \ref{app1} by Eqs. (\ref{F3res}),  (\ref{F2_2d}) and (\ref{F1fsi}). 
\bigskip

\section{Plane wave contribution}\label{pwc}

This contribution is displayed in the right panel in Fig. \ref{triangle}. 
According to the Feynman rules  it reads:
\begin{equation}\label{pw2}
J_{\mu,pw}=-\int \frac{(p+p'-2k)_{\mu}\Gamma_i\left(\frac{p}{2}-k,p\right)} {(p-k)^2-m^2+i\epsilon}\,\delta^{(4)}\left(k-p_s-\frac{p'}{2}\right)d^4k\,
\end{equation}
The delta-function follows from the four-momenta conservation in Fig. \ref{triangle}, right panel:
$$
\delta^{(4)}(k-p_1)=\delta^{(4)}\left(k-p_s-\frac{p'}{2}\right)
$$
We have introduced the total $p'$ and relative $p_s$ four-momentum  of the final (non-interacting) particles
\begin{eqnarray*}
2p_s &=&  p_1-p_2 \cr
p'      &=& p_1+p_2
\end{eqnarray*}

The spatial part   of $p_s$ in the rest frame of the final system $\vec{p'}=0$, determines the invariant final state mass $M'=2\sqrt{m^2+\vec{p}_s\,^2}$. 
One could calculate the integral over $d^4k$ by means of the delta-function. It is however interesting to keep this delta-function and carry out the integration later, 
once extracted the S-wave from the final state. 

Like in the case of FSI,  the form factor can be found by applying Eq.   (\ref{JF}) to the $J_{0,pw}$ component  of Eq. (\ref{pw2}), in the system of reference where $q_0=0$. That is:
\begin{equation}\label{Fpw}
F_{pw}=-\int \frac{(p_0-k_0)}{p_0}\;\frac{\Gamma_i\left(\frac{p}{2}-k,p\right)}{[(p-k)^2-m^2+i\epsilon]}
\int\frac{d\Omega_{\hat{p}_s}}{4\pi} 
\delta^{(4)}\left(k-p_s-\frac{p'}{2}\right)\,d^4k
\end{equation}

We have introduced here the additional integration over $\frac{d\Omega_{\hat{p}_s}}{4\pi}$  in the rest frame $\vec{p'}=0$ of the final state.
We remind that in the FSI contribution, calculated in the previous section, we decomposed the final state BS amplitude $\Gamma_f$ in  partial waves and took into account the S-wave only. The delta-function in (\ref{Fpw}) replaces now the final BS amplitude $\Gamma_f$.
Averaging this delta-function  over the solid angle $\hat{p}_s$ in the rest frame $\vec{p'}=0$ allows to select the partial S-wave in the plane wave. 
This is the meaning of the integral over $\frac{d\Omega_{\hat{p}_s}}{4\pi}$ in (\ref{Fpw}). 

The integration  over  $d\Omega_{\hat{p}_s}$ and part of the integration over $d^4k$  in (\ref{Fpw}) are done analytically in Appendix \ref{app_PW}.
The final result reads:
\begin{equation}\label{pw4}
F_{pw}=-\int_{k_-}^{k_+}\frac{(p_0-k_0)}{p_0}\; \Gamma_i \left(\frac{p}{2}-k,p\right)
\frac{1}{(p-k)^2-m^2+i\epsilon}\;
\frac{M'k}{2\varepsilon_k p_s p'}dk
\end{equation}
where the integration limits $k_{\mp}$ are defined in  (\ref{kmp}) and $p_s=\sqrt{{M'}^2/4-m^2}$. 
In  expression (\ref{pw4}) one must insert  $k_0=\sqrt{m^2+k^2}$ and, in the scalar product $k\cdot p=k_0p_0-zkp$ take the value  $z=z_0$ given by Eq. (\ref{z_0}). 
Variables $p_0$ and $p$ are the components of the initial four-momentum.
The value $p'$ is the spatial part
 of the total final state four-momentum in the frame  where $p'_0=p_0$. All these components are expressed 
in terms of the momentum transfer $Q^2$ in Eqs. (\ref{ppp}) and (\ref{p0}) from Appendix \ref{frame}.

Let us precise the arguments of $\Gamma_i \left(\frac{p}{2}-k,p\right)$ in (\ref{pw4}). 
They are defined analogously to the case of FSI, Eq. (\ref{G0}) in Appendix \ref{appB4}. 
Namely, solving  the BS equation in the rest frame $\vec{p}=0$, we find $\Gamma_i(k_0,|\vec{k}|)$ , where $k_0$ and $|\vec{k}|$  are also defined  in the rest frame $\vec{p}=0$. 
We should express them in the frame where $q_0=0$. 
These expressions are given in Appendix \ref{appB4}. 
That is, we have to  insert in (\ref{pw4}) the function $\Gamma_i(\tilde{k}_0,\vert\tilde{\vec{k}}\vert)$ with arguments  $\tilde{k}_0$ and $\vert\tilde{\vec{k}}\vert$ 
given by the first line of Eq. (\ref{arg}) from the Appendix  \ref{appB4}:
\begin{eqnarray*}
\tilde{k}_0&=&\frac{1}{2}M-\frac{1}{M}(k_0p_0-kpz),\\ 
\vert\tilde{\vec{k}}\vert&=&\sqrt{\frac{1}{M^2}(k_0 p_0-kpz)^2-k_0^2+\vec{k}^2}
\end{eqnarray*}
with $k_0=\sqrt{m^2+k^2}$ and $z_0$ defined in (\ref{z_0}).

\bigskip
To summarize these last  two sections we would like to emphasize that:
{\it (i)} the full PW contribution in the current is given by the simple equation  (\ref{pw2}),
{\it (ii)} the expression (\ref{pw4}) corresponds to the S-wave projection of the final  plane wave state,
and {\it (iii)} the full transition form factor -- including both the FSI and PW contributions -- is given by the sum (\ref{fffull}),
with  $F_{fsi}$ determined by Eq.  (\ref{FQ}) and $F_{pw}$  by (\ref{pw4}).

\section{Current conservation}\label{conserv}

As follows from Eq. (\ref{CC}), the conservation of the electromagnetic  current $q\cdot J=0$ implies $F'(Q^2)=0$ for any value of $Q^2$, that is $F'(Q^2)\equiv 0$.

To ensure this  conservation, all the contributions to the current, containing the interaction of a photon with a charged particle,  must be taken into account. 
In other words, in an interacting system, the true current is,  in general, not the free one.

\begin{figure}[h!]
\begin{center}
\mbox{\epsfxsize=8cm\epsffile{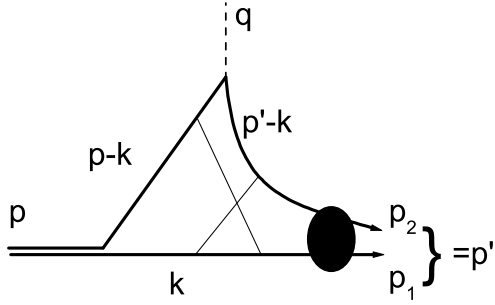}}
\end{center}
\caption{Cross-ladder contribution to the transition EM form factor. \label{cr_lad}}
\end{figure}

For example, for the kernel given by the sum of  ladder and cross-ladder, the full EM current should contain, in addition to the two (free) contributions displayed in  Fig. \ref{triangle}, 
the cross-ladder FSI contribution shown in Fig. \ref{cr_lad} and similar cross-ladder contribution for the plane wave (we suppose the exchanged particle to be neutral). 
The latter contributions are not free: they contain the  interaction of constituents. 
The sum of four contributions -- Fig. \ref{triangle} (left and right panels), Fig. \ref{cr_lad} (cross-ladder with FSI) and the corresponding cross-ladder PW
(not shown) -- must be conserved. 

In the case of the ladder kernel, the diagrams displayed in Fig. \ref{triangle}  provide the only contributions to the current. 
Therefore,  the current determined by these two graphs has to be conserved if the initial and final BS amplitudes are also obtained with the ladder kernel. 
At the same time, the expressions for the contributions (\ref{ffin0}) and (\ref{pw2}) to the current in terms of the BS amplitudes are universal -- they are the same for any BS amplitude (found with any kernel). The conservation of their sum is provided by  the particular properties of the BS  amplitudes determined  by the ladder kernel. The current is not conserved, if in Eqs.  (\ref{ffin0}) and (\ref{pw2}) one substitutes other BS amplitudes (not the ladder ones).
Therefore the current conservation (if any) provides a very strong test for the solutions themselves. 
In the this section we will calculate the form factor $F'(Q^2)$ and in the next section we will check numerically, whether it is identically zero or not. From (\ref{CC}) it follows that 
\begin{equation}\label{Fp}
F'(Q^2)=\frac{J\cdot q}{Q_c^2}
\end{equation}

In the expression (\ref{ffin0}) for $J_{fsi}$, after multiplying it by $q/Q_c^2$, we consider, for a moment, only the factor
\begin{equation}\label{factor0}
\frac{1}{Q_c^2}(p'-p)\cd(p+p'-2k)=\left.\frac{1}{Q_c^2}\left[{M'}^2-M^2-2(p'-p)\cd k\right]\right|_{p'_0=p_0}=\left(1-2\frac{\sqrt{Q^2}zk}{Q_c^2}\right)
\end{equation}

Instead of  the function $G(k_0,z,k)$ defined in Eq. (\ref{Gk0}) and given by Eq.  (\ref{G0}) in Appendix \ref{appB4}, we introduce the function:
\begin{equation}\label{Gp0}
G'(k_0,z,k)=\left(1-2\frac{\sqrt{Q^2}zk}{Q_c^2}\right) \Gamma_i(\tilde{k}_0,\tilde{k})\Gamma_f(\tilde{k'}_0,\tilde{k'}).
\end{equation}
$G'(k_0,z,k)$ is obtained from $G(k_0,z,k)$, replacing the factor  $(p_0-k_0)/p_0$ in $G(k_0,z,k)$  by the factor (\ref{factor0}):

\begin{equation}\label{factor1}
\frac{(p_0-k_0)}{p_0} \to \left(1-2\frac{\sqrt{Q^2}zk}{Q_c^2}\right),
\end{equation}
where as always  $z$  is the cosine of the angle between the integration variable $\vec{k}$ and the momentum $\vec{p}$ of the initial (bound state) system in the reference frame where $q_0=0$.

The FSI contribution $F'_{fsi}(Q^2)$ to the full form factor 
\begin{equation}\label{Fpp}
F'(Q^2)=F_{fsi}'(Q^2)+F_{pw}'(Q^2)
\end{equation}
is given by the same formulas than 
 for $F_{fsi}(Q^2)$, eq. (\ref{FQ}),  with the replacement $G(k_0,z,k) \to G'(k_0,z,k)$.

The PW contribution $F_{pw}'(Q^2)$ is calculated in a similar way. 
Namely $F_{pw}'(Q^2)$ is given by Eq. (\ref{pw4}) with  
the replacement (\ref{factor1}). The value $k_0$  is the same used in Eq. (\ref{pw4}) and  $z_0$ is defined in  Eq. (\ref{z_0}) from Appendix \ref{f2contrib}. 

In order to obtain $F'(Q^2)$  identically zero the two contributions FSI and PW to the full form factor $F'(Q^2)$  Eq. (\ref{Fpp})  must cancel each other.
Numerically this condition is never fulfilled exactly. The value of  $F'(Q^2)$ will be rather compared to $F(Q^2)$ and 
the current conservation would manifest itself in the fact that  $F'(Q^2)<<F(Q^2)$ for any value of $Q^2$.

\section{Numerical results}\label{numerics}

\begin{figure}[h!]
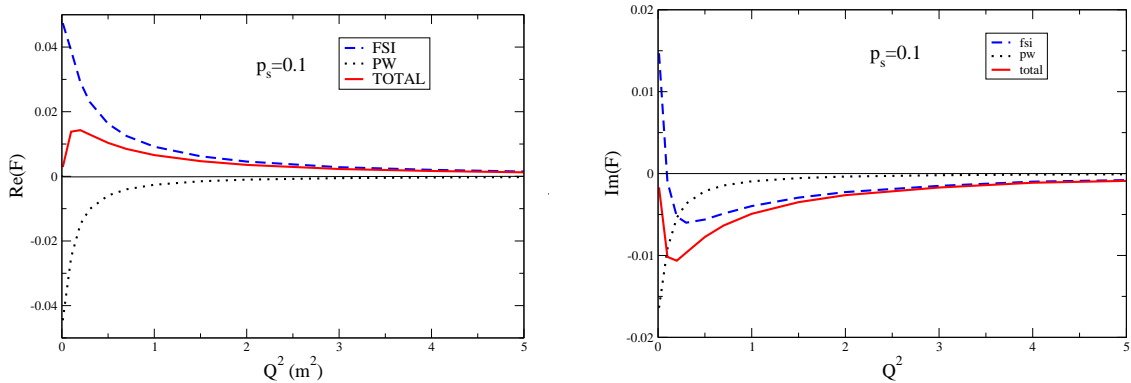

\begin{center}
\mbox{\epsfxsize=7.2cm
\epsffile{figure3a.eps}}
\hspace{0.5cm}
\mbox{\epsfxsize=7.cm
\epsffile{figure3b.eps}}
\caption{(color online). Transition EM form factor $F(Q^2)$ as a function of $Q^2$. Initial (bound) state corresponds to the binding energy $B=0.01\,m$; final (scattering) state corresponds to 
a relative momentum $p_s=0.1\,m$ (final state mass $M'=2.00998\,m$). FSI contribution is shown by the dashed curve and the PW one by a dotted curve. Full form factor  is shown by the solid curve. Left panel is the real part of form factor and right panel is the imaginary part.}
\label{ff}
\end{center}
\end{figure}
\vspace{0.5cm}

As an example, we have calculated the transition form factor for the  initial (bound) state  binding energy $B=0.01\,m$ (initial state mass $M=1.99\,m$) and for two values of the final (scattering) state  relative momentum $p_s=0.1\,m$ and $p_s=0.5\,m$ with corresponding final state masses $M'=2\sqrt{m^2+p_s^2}$ values $M'\approx 2.00998m$ and $M'\approx2.236m$. 

\vspace{.5cm}
\begin{figure}[h!]
\begin{center}
\mbox{\epsfxsize=7cm
\epsffile{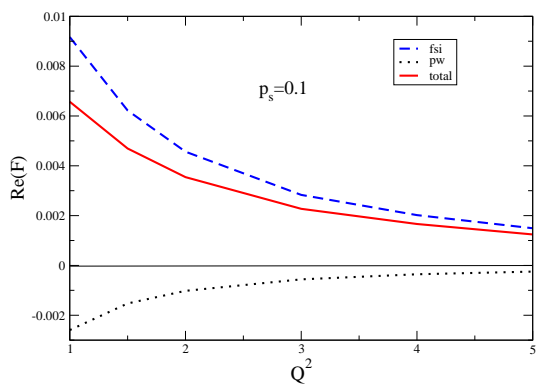}}
\caption{(color online). The same as on the left panel of Fig. \protect{\ref{ff}} for the tail of form factor $1\leq Q^2\leq 5$.}\label{ff-tail}
\end{center}
\end{figure}

In contrast to the elastic scattering, the inelastic transition form factor is complex. Its real and imaginary parts  as a function of $Q^2$ for $p_s=0.1\,m$ are shown in Fig. \ref{ff}, at left and right panels correspondingly.  One can see that at relatively small momentum transfer $Q^2 < 1$ both contributions -- FSI and PW -- are important and they considerably cancel each other.

The tail of the real part of form factor for $Q^2\geq 1$ and $p_s=0.1\,m$ is shown  in Fig. \ref{ff-tail}. 
In this momentum region, FSI dominates, especially when $Q^2$ increases. 

\vspace{.5cm}
\begin{figure}[h!]
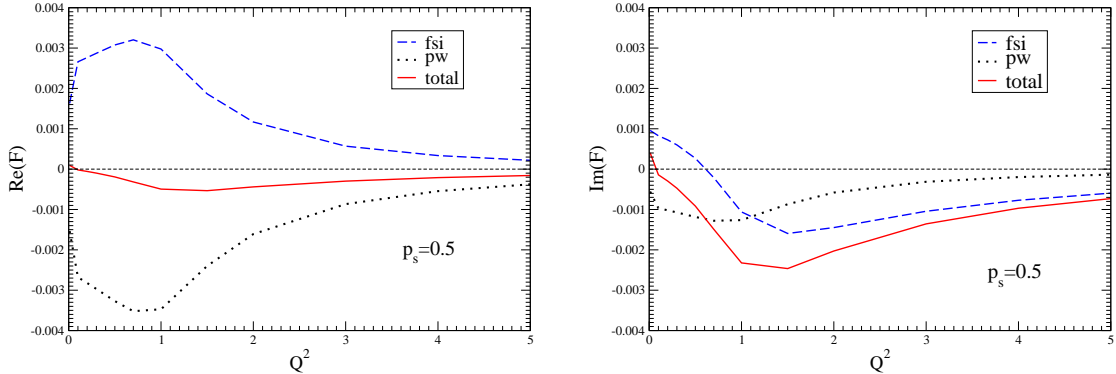

\begin{center}
\mbox{\epsfxsize=7cm
\epsffile{figure5a.eps}}
\hspace{0.5cm}
\mbox{\epsfxsize=7cm
\epsffile{figure5b.eps}}
\caption{(color online). The same as in Fig. \protect{\ref{ff}} for the final relative momentum $p_s=0.5\,m$ (final state mass: $M'=2.236\,m$).}\label{ff05}
\end{center}
\end{figure}

This is a natural behavior for the kinematics corresponding to Fig. \ref{ff-tail}.  
Indeed, due to small binding energy ($B=2m-M=0.01\,m$) the constituents in the initial state have small relative momentum. 
In the scattering process the photon transfers the large $Q^2$ value to one of the constituents only. 
 However, since their relative energy in the final state is also small ($M'-2m\approx 0.01\,m$), both constituents have also small relative momentum. 
 Therefore they move practically in the same direction, having both  large total momentum. 
 Since the second constituent does not interact with the photon, it can obtain a large total momentum only due to a strong interaction with  the first constituent. 
This explains why in this kinematics the final state interaction (re-scattering) determines the tail of the form factor and dominates over the plane wave.
\vspace{0.5cm}

\begin{figure}[htbp]
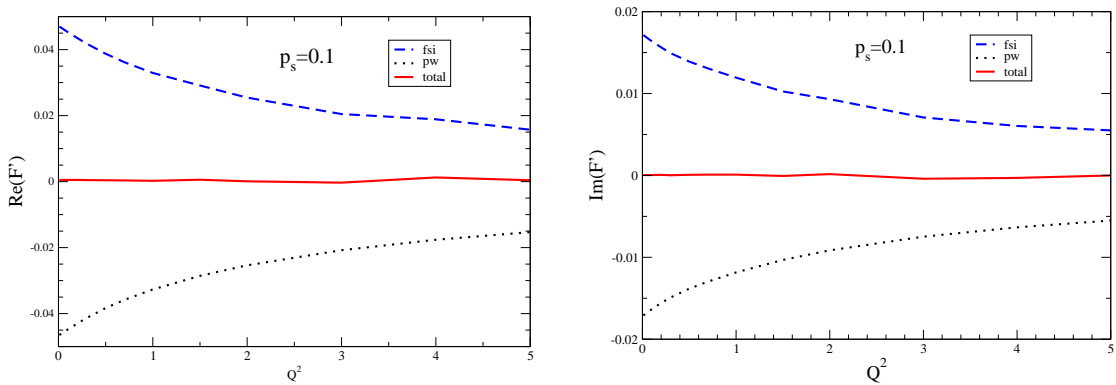

\begin{center}
\mbox{\epsfxsize=7cm
\epsffile{figure6a.eps}}
\hspace{0.5cm}
\mbox{\epsfxsize=7cm
\epsffile{figure6b.eps}}
\caption{(color online). Transition EM form factor $F'(Q^2)$ as a function of $Q^2$  for $p_s=0.1$. Other parameters and notations are the same as in Fig. \protect{\ref{ff}}.}
\label{figFp}
\end{center}
\end{figure}

\bigskip
The transition EM form factor for  larger final  relative momentum $p_s=0.5\,m$, final state mass $M'=2.336\,m$, and the same values of other parameters is shown in Fig. \ref{ff05}. 
For this larger value of the final state mass, the FSI contribution is still significant, but it does not dominate anymore.
In the real part, FSI and PW  contributions considerably cancel each other. 
\vspace{0.5cm}

\begin{figure}[htbp]
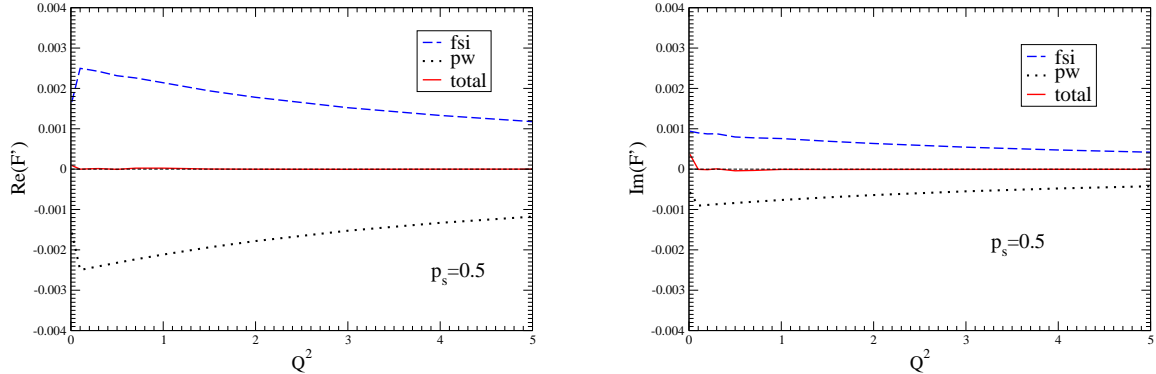

\begin{center}
\mbox{\epsfxsize=7cm
\epsffile{figure7a.eps}}
\hspace{1.cm}
\mbox{\epsfxsize=7cm
\epsffile{figure7b.eps}}
\caption{(color online). Transition EM form factor $F'(Q^2)$ as a function of $Q^2$  for $p_s=0.5$. Other parameters and notations are the same as in Fig. \protect{\ref{ff}}.}
\label{figFp05}
\end{center}
\end{figure}

As mentioned above, the form factor $F'(Q^2)$ -- which vanishes if the current is conserved -- 
is obtained from $F(Q^2)$ by the replacement (\ref{factor1}) in the integrand. 
The corresponding numerical results   for $p_s=0.1$ and $p_s=0.5$ are shown in Figs. \ref{figFp} and \ref{figFp05} respectively. 
We see that, in comparison with the $F(Q^2)$ results of Figs. \ref{ff}-\ref{ff05}, the value of $F'(Q^2)$ is indistinguishable from zero. 
This very small value is a result of a cancellation between FSI and PW contributions. 
We conclude that in the model considered, the current is conserved. 

\bigskip\bigskip
\begin{figure}[h!]
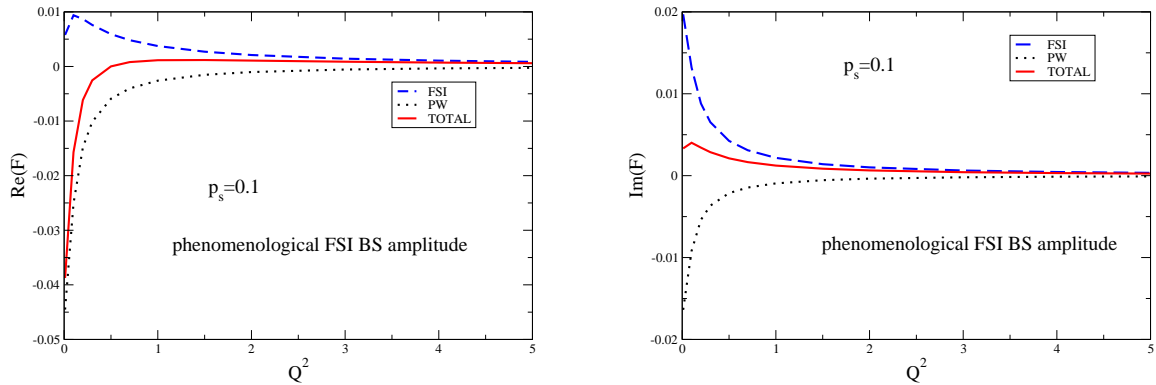

\begin{minipage}[h!]{16cm}
\begin{center}
\mbox{\epsfxsize=7cm
\epsffile{figure8a.eps}}
\hspace{1.cm}
\mbox{\epsfxsize=7cm
\epsffile{figure8b.eps}}
\caption{(color online). Transition EM form factor $F(Q^2)$ as a function of $Q^2$  for $p_s=0.1$ calculated with the "phenomenological" FSI function 
(\ref{G_phenom}).  The notations are the same as in Fig. \protect{\ref{ff}}.}
\label{F_phenom}
\end{center}
\end{minipage}
\end{figure} 

\bigskip
Though the current conservation is natural, the cancellation of FSI and PW contributions is rather delicate and provides as a strong test
of a calculation. Indeed, 
both FSI and PW contributions contain the same initial bound state BS amplitude. At the same time, FSI contribution contains the scattering state BS amplitude, 
whereas the PW contribution -- does not. Their cancellation takes place provided both BS amplitudes -- the bound and the scattering one -- as well as the current operator are consistent with each other, i.e.  if they are correctly found in the same dynamics.

To illustrate how a violation of this consistency would affect the current conservation, we replaced the final BS amplitude, found for the OBE kernel, by an "ad-hoc" function 
\begin{equation}\label{G_phenom}
\Gamma_f(k_0,z,k)= \frac{1}{(k_0^2+a^2)(k^2+b^2)}
\end{equation}
without changing the initial BS amplitude and the current. The transition form factor $F(Q^2)$ calculated with the function (\ref{G_phenom}) for $a^2=1.5$, $b^2=1$ is shown in Fig. \ref{F_phenom}. 
Apparently it has a typical behavior and nothing indicates that it is a wrong result.

\begin{figure}[htbp]
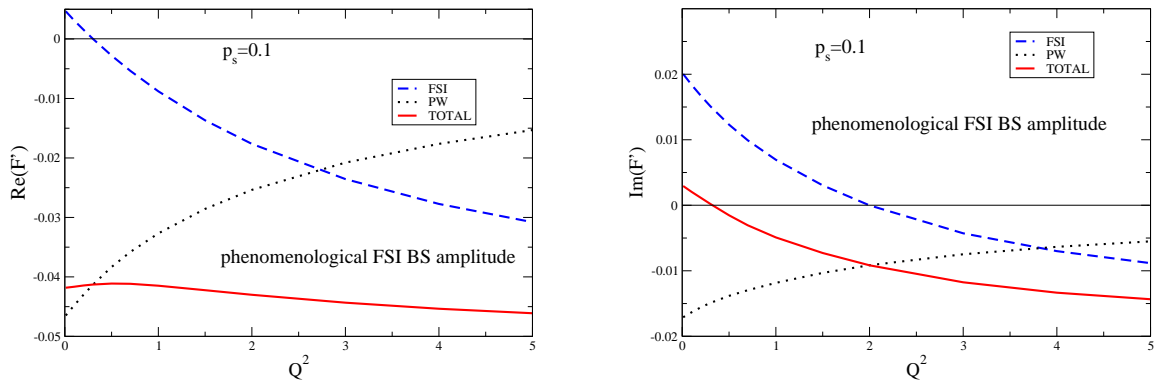

\begin{center}
\mbox{\epsfxsize=7cm
\epsffile{figure9a.eps}}
\hspace{1.cm}
\mbox{\epsfxsize=7cm
\epsffile{figure9b.eps}}
\caption{(color online). Transition EM form factor $F'(Q^2)$ as a function of $Q^2$  for $p_s=0.1$ calculated with the "phenomenological" FSI function 
(\ref{G_phenom}).  The notations are the same as in Fig. \protect{\ref{ff}}.}\label{Fp_phenom}
\end{center}
\end{figure}

To see that  we have displayed  in Fig. \ref{Fp_phenom} the transition form factor $F'(Q^2)$ calculated with the same function (\ref{G_phenom}). 
We see in this figure that $F'$  is different from zero and   of the same order  than form factor $F(Q^2)$.
This means that the EM current  calculated with the "phenomenological" FSI function  (\ref{G_phenom}) is not correct and therefore  the form factor $F(Q^2)$
extracted from this current -- shown in Fig. \ref{F_phenom} -- is also incorrect.

\bigskip
The study of the numerical stability as a function of  the number  of Gaussian integration points $n_G$, shows that 
the sum $F'_{tot}=F'_{FSI}+F'_{PW}$ decreases with  $n_G$.  
For $n_G=64$ it  is two order of magnitude smaller than each  $F'_{FSI}$ and $F'_{PW}$ taken separately and also than the form factor $F$. 
This means that there exist a cancellation between   $F'_{FSI}$ and $F'_{PW}$, and hence the current conservation, with  a numerical precision of about 1\%.  
Increasing the value of $n_G$ from 64 to 128 does not improve the result (does not reduce $F'_{tot}$). 
Apparently,  the precision of $F'_{tot}$ is determined by the accuracy  of the numeric  solutions for the initial and final BS amplitudes.

The numerical results for $F(Q^2)$ and $F'(Q^2)$, calculated with $n_G=64$, $p_s=0.1$  are given in the Table \ref{tab3}.
For $Q^2=1.5$ the value of $F'_{tot}$ is by one order of magnitude smaller than $F$ and two orders of magnitude smaller than  $F'_{FSI}$ and $F'_{PW}$). 
When $Q^2$ increases up to $Q^2=5$, the cancellation becomes worser and almost disappears, though
$F'_{tot}$ is still a few times smaller than $F'_{FSI}$ and  $F'_{PW}$. 
This  is related to the fact that the BS amplitudes  was computed in a finite domain of  variables $k_0,k$ which,  at large values of $Q^2$,  
is not enough to ensure enough accurate result for the BS solution. 

\begin{table}[htbp]
\begin{center}
\caption{\mbox{The value $F'(Q^2)$ for FSI, PW and tot=FSI+PW vs. $Q^2$ in comparison to $F(Q^2)$;
$p_s=0.1.$}}\label{tab3}
\begin{tabular}{cc ccc}\hline
$Q^2$       \phantom{0}&\phantom{0} $F(Q^2)$ \phantom{0}&\phantom{0}  $F'_{FSI}(Q^2)$     \phantom{0}&\phantom{0}  $F'_{PW}(Q^2)$  \phantom{0}&\phantom{0} $F'_{tot}(Q^2)$  \\\hline
0.01  \phantom{0}&\phantom{0}   $2.948\mbox{-03} -1.571\mbox{-03}$  \phantom{0}&\phantom{0} $4.703\mbox{-02}+i1.732\mbox{-02}$ \phantom{0}&\phantom{0} $-4.648\mbox{-02} -i1.709\mbox{-02}$ \phantom{0}&\phantom{0}
$5.490\mbox{-04} +i2.304\mbox{-04}$ \cr
0.1 \phantom{0}&\phantom{0} $1.391\mbox{-02} -i1.008\mbox{-02}$  \phantom{0}&\phantom{0}  $4.541\mbox{-02}+i1.641\mbox{-02} $ \phantom{0}&\phantom{0} $-4.478\mbox{-02} -i1.633\mbox{-02}$   \phantom{0}&\phantom{0}
$6.268\mbox{-04} +i8.139\mbox{-05}$ \cr
0.5  \phantom{0}&\phantom{0}  $1.047\mbox{-02} -7.653\mbox{-03}$ \phantom{0}&\phantom{0} $3.904\mbox{-02}+i1.401\mbox{-02}$  \phantom{0}&\phantom{0} $-3.833\mbox{-02} -i1.383\mbox{-02}$ \phantom{0}&\phantom{0} 
$7.089\mbox{-04} +i1.848\mbox{-04}$ \cr
1. \phantom{0}&\phantom{0} $6.640\mbox{-03} -i4.841\mbox{-03}$ \phantom{0}&\phantom{0} $3.321\mbox{-02}+i1.181\mbox{-02}$  \phantom{0}&\phantom{0}  $-3.268\mbox{-02} -i1.184\mbox{-02}$ \phantom{0}&\phantom{0}
$5.303\mbox{-04} -i3.634\mbox{-05}$ \cr
2. \phantom{0}&\phantom{0} $3.573\mbox{-03}- i2.460\mbox{-03}$  \phantom{0}&\phantom{0} $   2.639\mbox{-02}  +i8.478\mbox{-03}    $ \phantom{0}&\phantom{0} $    -2.537\mbox{-02} -i9.150\mbox{-03}     $ \phantom{0}&\phantom{0} $ 1.018\mbox{-03}  -i6.726\mbox{-04}  $  \cr
3.  \phantom{0}&\phantom{0} $2.362\mbox{-03}- i1.589\mbox{-03}$  \phantom{0}&\phantom{0} $    2.234\mbox{-02}  +i7.062\mbox{-03}     $ \phantom{0}&\phantom{0} $    -2.079\mbox{-02} -i7.482\mbox{-02}      $ \phantom{0}&\phantom{0} $  1.552\mbox{-03} -i4.204\mbox{-04} $ \cr
4.  \phantom{0}&\phantom{0}  $1.718\mbox{-03}- i1.109\mbox{-03}$  \phantom{0}&\phantom{0} $    1.897\mbox{-02}  +i6.511\mbox{-03}     $ \phantom{0}&\phantom{0} $    -1.763\mbox{-02} -i6.339\mbox{-02}      $ \phantom{0}&\phantom{0} $  1.341\mbox{-03} +i1.721\mbox{-04} $ \cr
5.  \phantom{0}&\phantom{0} $1.446\mbox{-03}- i8.313\mbox{-04}$  \phantom{0}&\phantom{0} $    1.993\mbox{-02}  +i6.093\mbox{-03}     $ \phantom{0}&\phantom{0} $    -1.531\mbox{-02} -i5.506\mbox{-02}      $ \phantom{0}&\phantom{0} $  4.614\mbox{-03} +i5.873\mbox{-04} $ 
 \\ \hline
\end{tabular}
\end{center}
\end{table}

\section{Conclusion} \label{concl}

We have presented the first results of the transition electromagnetic form factor for the electrodisintegration of a two-body bound system described by  the 
Bethe-Salpeter equation in Minkowski space.
Calculations have been performed in a self-consistent way. 
The initial (bound state) and final (scattering state) BS amplitudes were found by solving the equation with  the method developed in our previous works \cite{CK_PLB_PRD}
and an OBE kernel. 

We have shown that,  provided the bound and scattering state Bethe-Salpeter amplitudes as well as the  operator of EM current are consistent
with each other, the electromagnetic current is conserved. 
If this consistency is destroyed, the conservation  is violated.  
This violation has two consequences.

First, the decomposition of the  current in form factors obtains an additional contribution (second term in Eq. (\ref{ffc})) which does not satisfy the equality $J\cdot q=0$. 
However, the appearance of this term itself does not make any influence on observables -- it does not contribute in the scattering amplitude --
due to conservation of the electromagnetic current of the incident electron.

Second, and most important, is the fact that the non-conservation of the calculated  current makes it physically meaningless. 
One cannot extract from a deficient current a reliable transition form factor. 
It is thus mandatory, in practical calculations, like e.g. in the deuteron electrodisintegration, to check the current conservation. 
If the form factor, responsible for non-conservation of the  current, turns out to be comparable with the physical ones,  one can hardly trust  the calculated physical form factors too. 

The widely used recipe, consisting in replacing the non-conserved current $J_{\mu}$  
by the conserved combination $\tilde{J}_{\mu}= J_{\mu}-q_{\mu}(J\cdot q)/q^2$ hides the problem but does not solve it. 
This combination $\tilde{J}_{\mu}$ satisfies tautologically the  current conservation for any $J_{\mu}$, not only  for 
the correct one and thus offers no any guarantee  to the result. 
With an incorrect current, one cannot find the correct transition form factor, neither from $J_{\mu}$ nor from $\tilde{J}_{\mu}$. 

The current conservation appears as a numerically subtle phenomenon since it manifests itself as a cancellation of large contributions: FSI and PW. 
To see it unambiguously, the solution of the BS equation should be found with high  enough precision. 

\appendix

\section{Kinematics}\label{frame}

As mentioned in Sec. \ref{tff}, it is convenient to carry out these calculations in the system of reference where $p'_0=p_0$, i.e. $q_0=0$. 
In the elastic case it coincides with the Breit frame $\vec{p}+\vec{p'}=0$ and $|\vec{p}|=|\vec{p'}|$,  $p'_0=p_0$. 

This system exists in the inelastic case $M'\neq M$ too. Indeed, one can easy check that in the reaction $e+d\to e'+(np)$ the momentum transfer $q^2=(p-p')^2=(p_0-p'_0)^2-(\vec{p}-\vec{p'})^2$ 
is always negative. In particular, its maximal value (reached at the minimal value of $s=(M'+m_e)^2$, with $m_e$ the electron mass) is still negative:
$$
q^2\leq -\frac{m_e({M'}^2-M^2)}{M'+m_e}<0.
$$
Therefore one can find a reference frame where this negative value contains only spatial components: $q^2=(p-p')^2=-(\vec{p}-\vec{p'})^2$ so that $p'_0=p_0$. Moving this frame, without changing the value $q_0=0$,
one can also make the vectors $\vec{p}$ and $\vec{p'}$  collinear, i.e. either parallel or anti-parallel, so that  $Q^2=-q^2=(\vec{p}-\vec{p'})^2$  is equal either to $(p-p')^2$ (if $\vec{p}$ and $\vec{p'}$ are parallel) or $(p+p')^2$ (if $\vec{p}$ and $\vec{p'}$ are anti-parallel),
depending on the $Q^2$ value. We  denote 
$p=\mid\vec{p}\mid$ and $p'=\mid\vec{p'}\mid$. 
We will precise below  when one should take the  plus or minus sign. 

In case of an elastic collision $M'=M$, in the Breit frame with $q_0=0$ (i.e. $p'_0=p_0$) and with the initial and final momenta satisfying  $\vec{p}+\vec{p'}=0$ and $|\vec{p'}|=|\vec{p}|$, 
the scattered system  moves in the opposite direction than the incoming one and 
 this is the only possibility to get a non-zero momentum transfer.  
 
In the inelastic case $M'>M$, still in the reference frame with  $p'_0=p_0$ and with  the  collinear momenta $\vec{p}$, $\vec{p'}$  (now with $|\vec{p'}|\neq|\vec{p}|$ but $|\vec{p'}|<|\vec{p}|$), there exists a critical momentum transfer $Q_c^2$.  If the momentum transfer $Q^2$ is smaller that $Q_c^2$, it is not enough to change the direction of initial momentum $\vec{p}$ into the opposite one. 
In this kinematics, the final momentum $\vec{p'}$ after collision  remains  parallel to $\vec{p}$, though with $p'$  smaller than $p$. 
However, for $Q^2>Q_c^2$ the final momentum $\vec{p'}$ changes its direction relative to $\vec{p}$ like in the elastic collision. 
That is, when $Q^2$ increases, the final momentum $\vec{p'}$, being first parallel to $\vec{p}$, vanishes 
and  appears again in a direction opposite to $\vec{p}$. When it crosses zero $\vec{p'}=0$
(provided $p'_0=p_0$), the corresponding  momentum transfer is $Q_c^2=p^2$. 
We get in this case:
\begin{eqnarray*}
 p'_0=p_0 &\to&  \sqrt{{p'}^2+{M'}^2}=\sqrt{{p}^2+{M}^2}
 \\
 &\to&  M'=\sqrt{{p}^2+{M}^2}.
\end{eqnarray*}
From last equality we find the critical value  $Q^2_c=p^2$  for which $p'=0$
\begin{equation}\label{Qc2}
Q_c^2={M'}^2-M^2.
\end{equation}
In the  frame where we perform the calculations ($q_0=0$), the following kinematical relations holds:
\begin{eqnarray}
(\vec{p'}- \vec{p})^2  &=& (p'-\sigma p)^2                       \label{pp} \\
\vec{k}\cdot\vec{p'}  &=& \sigma \; k p' z                      \label{k_dot_pp}  \\
(\vec{p}- \vec{k})^2  &=&  p^2-2zpk+k^2                       \label{pmk2}  \\
(\vec{p'}- \vec{k})^2 &=& {p'}^2-2\sigma zp'k+k^2           \label{ppmk2} \\
 \sqrt{Q^2}              &=& p-\sigma p'                                  
\end{eqnarray}
where we have introduced  the "sign" variable $\sigma$ depending on $Q^2$ and $Q^2_c$:
\begin{equation}\label{signe}
\sigma(Q^2,Q^2_c) =\left\{   
\begin{array}{lcl}   
+1   & if &  Q^2<Q^2_c\cr   
- 1   & if &  Q^2>Q^2_c 
 \end{array}  \right. 
\end{equation} 
and denote hereafter (abusively) : $k=\mid\vec{k}\mid$.

From the requirement $\sqrt{{M'}^2+{p'}^2}= \sqrt{M^2+p^2}$ we find relation between $p$ and $p'$:
\begin{eqnarray*}
p'  &=&  \sqrt{{M}^2-{M'}^2+{p}^2} \cr 
p   &=& \sqrt{{M'}^2-M^2+{p'}^2}
\end{eqnarray*}
and so:
\begin{eqnarray}
p'&=&\frac{|  Q_c^2-Q^2|}{2\sqrt{Q^2}}  \cr
p &=&\frac{   Q_c^2+Q^2 }{2\sqrt{Q^2}}   \label{ppp} \\
p_0=p'_0&=&\frac{\sqrt{((M'-M)^2+Q^2)((M'+M)^2+Q^2)}}{2\sqrt{Q^2}} \label{p0}
\end{eqnarray}

\section{Calculating the FSI contribution to the transition form factor}\label{app1}

The  contribution of FSI to the transition form factor is given in Eq. (\ref{FQ}) as a sum of three terms $F_{3,2,1}$.
In their turn, $F_{3,2,1}$ are obtained by integrating  over $dk_0d^3k$  the three functions $f_{3,2,1}$ defined in (\ref{f3}-\ref{f1}). 
We detail in what follows the calculation of these three contributions.
\bigskip

\subsection{$F_3$ contribution}\label{F_3}

Let us first consider the $F_3$ contribution
\[  F_3(Q^2)=\frac{i}{(2\pi)^3}  \; \int_0^{\infty}  k^2dk \int_{-1}^{+1} dz\int_{-\infty}^{+\infty} dk_0 \;  f_3(k_0,z,k) \]
with   $f_{3}$  given by Eq. (\ref{f3}). 

As a function of $k_0$,  $f_{3}$  contains six poles in $k_0$-variable at the points
\begin{eqnarray*}
k_0  &=&          +  \varepsilon_{\vec{k}} \cr
k_0  &=&           - \varepsilon_{\vec{k}}  \cr
k_0  &=&   p_0  + \varepsilon_{\vec{p}-\vec{k}}   \cr
k_0  &=&   p_0  -  \varepsilon_{\vec{p}-\vec{k}}   \cr
k_0  &=&   p_0  + \varepsilon_{\vec{p'}-\vec{k}}    \cr
k_0  &=&   p_0  -  \varepsilon_{\vec{p'}-\vec{k}} 
\end{eqnarray*}
and eventually other singularities resulting from $G(k_0,z,k)$. 
We subtract and add  to $f_3$ a function $h_3$ depending on the same variables and  having  poles only in the variable $k_0$. That is
\begin{equation}\label{f3-h3+h3}
f_{3}=f_3-h_3+h_3 = \bar{f}_3 + h_3
\end{equation}
with
\begin{equation}\label{h3}  
h_3(k_0,z,k) =\frac{g_1(z,k)}{k_0-\varepsilon_{\vec{k}}}+ \frac{g_2(z,k)}{k_0+\varepsilon_{\vec{k}}}+ \frac{g_3(z,k)}{k_0-p_0-\varepsilon_{\vec{p}-\vec{k}}} + \frac{g_4(z,k)}{k_0-p_0+\varepsilon_{\vec{p}-\vec{k}}}
+ \frac{g_5(z,k)}{k_0-p'_0-\varepsilon_{\vec{p'}-\vec{k}} }+ \frac{g_6(z,k)}{k_0-p'_0+\varepsilon_{\vec{p'}-\vec{k}}}
\end {equation}
The coefficients $g_{i} $, independent on $k_0$,  are determined by imposing that the difference
\begin{equation} 
\bar{f}_3\equiv f_3-h_3 
\end{equation}
is regular  in $k_0$. For instance
\[ \lim_{k_0\to \varepsilon_k } (k_0-\varepsilon_k  )\bar{f}_3= 0  \qquad \Longleftrightarrow\qquad     \lim_{k_0\to \varepsilon_k } (k_0-\varepsilon_k  )f_3 =  g_1(z,k),  \]
and similarly for other poles.

This gives
\begin{eqnarray}
g_1(z,k) &=& +\;  \frac{G(+\varepsilon_k,z,k)}               { 2\varepsilon_k  [ (p_0-\varepsilon_k )^2 - \epsilon^2_{p-k} ] \; [ (p_0-\varepsilon_k )^2 - \epsilon^2_{p'-k} ]  }     \cr 
g_2(z,k) &=& -\;  \frac{G(-\varepsilon_k,z,k)}                 { 2\varepsilon_k  [ (p_0+\varepsilon_k )^2 - \epsilon^2_{p-k} ] \; [ (p_0+\varepsilon_k )^2 - \epsilon^2_{p'-k} ]  }   \cr
g_3(z,k) &=& +\;  \frac{G(p_0+\varepsilon_{p-k} ,z,k)} { 2\varepsilon_{p-k} [ (p_0+\varepsilon_{p-k})^2 -\varepsilon_k^2] \; (\varepsilon_{p-k} ^2-  \varepsilon_{p'-k} ^2)     }         \cr
g_4(z,k) &=& -\;   \frac{G(p_0-\varepsilon_{p-k} ,z,k)}  { 2\varepsilon_{p-k}  [ (p_0-\varepsilon_{p-k})^2 -\varepsilon_k^2] \; (\varepsilon_{p-k} ^2-  \varepsilon_{p'-k} ^2)  }            \cr
g_5(z,k) &=& -\;   \frac{G(p_0+\varepsilon_{p'-k} ,z,k)} { 2\varepsilon_{p'-k}[ (p_0+\varepsilon_{p'-k})^2 -\varepsilon_k^2] \; (\varepsilon_{p-k} ^2-  \varepsilon_{p'-k} ^2)   }                      \cr
g_6(z,k) &=& +\;  \frac{G(p_0-\varepsilon_{p'-k} ,z,k)} { 2\varepsilon_{p'-k} [ (p_0-\varepsilon_{p'-k})^2 -\varepsilon_k^2] \; (\varepsilon_{p-k} ^2-  \varepsilon_{p'-k} ^2)   }              \label{g_i}
\end{eqnarray}
and the function $\bar{f}_3$ obtains the form:
\begin{eqnarray}
\bar{f}_3(k_0,z,k)\hspace{-0.2cm}
&=&\hspace{-0.2cm}   \frac{G(k_0,z,k)}{  [k_0-\varepsilon_k] \; [k_0+\varepsilon_k]  \;  [  k_0-p_0-\varepsilon_{p-k} ]   \;  [  k_0-p_0 +\varepsilon_{p-k} ]   \;  [  k_0-p_0 -\varepsilon_{p'-k} ]   \;  [  k_0-p_0 +\varepsilon_{p'-k} ] }  \label{f3bar} \\
&-& \hspace{-0.2cm}   \frac{g_1(z,k)}{k_0-\varepsilon_k } 
-  \frac{g_2(z,k)}{k_0+\varepsilon_k } - \frac{g_3(z,k)}{k_0-p_0-\varepsilon_{p-k} } - \frac{g_4(z,k)}{k_0-p_0+\varepsilon_{p-k} }  - \frac{g_5(z,k)}{k_0-p_0-\varepsilon_{p'-k }} - \frac{g_6(z,k)}{k_0-p_0+\varepsilon_{p'-k} }  \nonumber
\end{eqnarray}

\bigskip
The PV integral over $k_0$ of the remaining integrand $h_3$ in (\ref{f3-h3+h3}), vanish in the  full integration domain:
\begin{equation}\label{FVC}
\mbox{PV}\int_{-\infty}^{\infty} h_3\,dk_0=0.
\end{equation}
However, in the numerical solution we restrict the integration domain to a finite interval $k_0\in[-L,+L]$. The integral (\ref{FVC}) is no longer zero and finite volume correction must be  taken into account.
The integral over the finite domain $[-L,+L]$ of the function $h_3$ given in (\ref{h3}) is analytic and  reads:
\begin{eqnarray}
f_{3,fv}(z,k) \equiv \mbox{PV}\int_{-L}^{+L} dk_0 \; h_3 (k_0,k,z) &=& g_1(k,z) \log{  \left| \frac{L-\varepsilon_kÃ}{L+\varepsilon_k}  \right|   }  \hspace{1.2cm}  +  g_2(k,z) \log{  \left| \frac{L+\varepsilon_kÃ}{L-\varepsilon_k}  \right|   }  \cr
                                                             &+& g_3(k,z) \log{  \left| \frac{L-p_0-\varepsilon_{p-k} Ã} {L+p_0+\varepsilon_{p-k} }  \right|   }                             +  g_4(k,z) \log{  \left| \frac{L-p_0+\varepsilon_{p-k} } {L+p_0-\varepsilon_{p-k} }  \right|}  \cr
                                                             &+& g_5(k,z) \log{  \left| \frac{L-p_0-\varepsilon_{p'-k}  } {L+p_0+\varepsilon_{p'-k} }  \right|   }                            + g_6(k,z) \log{  \left| \frac{L-p_0+\varepsilon_{p'-k} } {L+p_0-\varepsilon_{p'-k} }  \right|}  \label{f3fv}
\end{eqnarray}
For the raisons that will become clear latter, we will include the above finite volume contributions 
\begin{equation}\label{F3fv}
F_{3,fv}(Q^2)= \frac{i}{(2\pi)^3}\int_0^{\infty}k^2dk \int_{-1}^1 dz\; f_{3,fv}(z,k) 
\end{equation}
in the $F_2$ contribution, to be discussed in the next section.

\bigskip
The  $F_3(Q^2)$ will thus be given by the following three-dimensional integral:
\begin{equation}\label{F3res}
F_{3}(Q^2)=\frac{i}{(2\pi)^3}\int_0^{\infty}k^2dk  \int_{-1}^1 dz \int_{-L}^Ldk_0\; \bar{f}_3(k_0,z,k) 
\end{equation}
with  $\bar{f}_3(k_0,z,k)$ defined in (\ref{f3bar}).

The integrand in (\ref{F3res}) is by construction a smooth function in $k_0$.
Concerning the integration over the variables $k$ and $z$, a further inspection is needed.
For $M'>2m$ the value $p_0-\varepsilon_{\vec{k}}-\varepsilon_{\vec{p'}-\vec{k}}=p'_0-\varepsilon_{\vec{k}}-\varepsilon_{\vec{p'}-\vec{k}}$ in  the denominator  of (\ref{f3bar}) can vanish.
Indeed, a state with mass $M'>2m$ can decay in two particles with masses $m$, and  in the moving frame this implies $p'_0=\varepsilon_{\vec{k}}+\varepsilon_{\vec{p'}-\vec{k}}$. 
The $g_1$ and $g_6$ terms  in Eq. (\ref{f3bar}) are therefore singular in variable $\vec{k}$. 
These two singularities cancel each other since an expansion of $\bar{f}_3$ in the variable $\varepsilon_{\vec{k}}$ near 
$\varepsilon_{\vec{k}}=p_0-\varepsilon_{\vec{p'}-\vec{k}}$  shows that the result is proportional to  $\sim {\cal O}((\varepsilon_{\vec{p'}-\vec{k}}-p'_0)^0)$.
As a consequence, the integrand $\bar{f}_3$ has no additional singularities in $k,z$ and
can be safely integrated  numerically by standard methods.

\subsection{$F_2$ contribution}\label{f2contrib}

Let us now calculate the contribution
\begin{equation}\label{F21}
F_2(Q^2)=\frac{i}{(2\pi)^3}\; \int_0^{\infty}  k^2dk \int_{-1}^{+1} dz\int_{-\infty}^{+\infty} dk_0 \;   \; f_2  (k_0,z,k)     
\end{equation}
with $f_{2}$  given by Eq. (\ref{f2}).

The  integration  over $k_0$-variable can be performed analytically by means of the delta-functions. The result -- denoted  $\tilde{f}_2$ -- is expressed  in terms of functions  $g_i$ defined in (\ref{g_i}) and reads:
\begin{equation}\label{f_2}
\tilde{f}_2(z,k)=\int_{-\infty}^{+\infty}dk_0 \; f_2(k_0,z,k)  
= -i\pi  \Big\lbrace{g_1(z,k)-  g_2(z,k) + g_3(z,k) -  g_4(z,k) -g_5(z,k)- g_6(z,k)}\Big\rbrace
\end{equation}
As one can see, $\tilde{f}_2(z,k)$ has a similar structure and depends on the same variables than the finite volume corrections $ f_{3,fv}(z,k) $ described above in (\ref{f3fv}).
It is thus natural to include both contributions in the same integrand
\[ \bar{f}_2 (z,k) = \tilde{f}_2(z,k) +f_{3,fv}(z,k) =  \sum_{i=1}^6 c_i(k) g_i(k)  \]
by introducing the coefficients 
\begin{equation}\label{c_i}
\begin{array}{lclcl}
c_1(k) &=&     -i\pi   &+&  \log{  \left| \frac{L-\varepsilon_kÃ}{L+\varepsilon_k}  \right|   }   \cr
c_2(k) &= &  +i\pi    &+&   \log{  \left| \frac{L+\varepsilon_kÃ}{L-\varepsilon_k}  \right|   }   \cr
c_3(k) &=&   -i\pi     &+&  \log{  \left| \frac{L-p_0-\varepsilon_{p-k} Ã} {L+p_0+\varepsilon_{p-k} }  \right| }  \cr
c_4(k) &=&  +i\pi     &+&   \log{  \left| \frac{L-p_0+\varepsilon_{p-k} } {L+p_0-\varepsilon_{p-k} }  \right| }       \cr
c_5(k) &=&  +i\pi     &+&  \log{  \left| \frac{L-p_0-\varepsilon_{p'-k}  } {L+p_0+\varepsilon_{p'-k} } \right| }   \cr  
c_6(k) &=&  +i\pi     &+&  \log{  \left| \frac{L-p_0+\varepsilon_{p'-k} } {L+p_0-\varepsilon_{p'-k} }  \right| }   
\end{array}
\end{equation}
The $F_2$  contribution is then given by the two-dimensional integral 
\begin{equation}\label{F2_2d}
 F_2(Q^2)=\frac{i}{(2\pi)^3}\; \int_0^{\infty}  k^2dk \int_{-1}^{+1} dz  \; \bar{f}_2  (z,k)       
\end{equation} 
The integral (\ref{F2_2d}) over the $z$ variable requires some care since both $g_1$ and $g_6$ can have  pole singularities in $z$. In contrast to the 
function $\bar{f}_3$, Eq. (\ref{f3bar}), these singularities  in $\bar{f}_2$ do not cancel each other.  However, they can be integrated analytically over $z$, so the pole singularities turn into the log-ones.

\bigskip
Let us first consider the $g_1$ term:
\[  g_1(z,k)=  \frac{G(+\varepsilon_k,z,k)}               { 2\varepsilon_k  [ (p_0-\varepsilon_k )^2 - \epsilon^2_{p-k} ] \; [ (p_0-\varepsilon_k )^2 - \epsilon^2_{p'-k} ]  } \]
The denominator vanishes if
 \[  (p_0-\varepsilon_k)^2- \epsilon^2_{p'-k}=0  \quad\Longleftrightarrow\quad  2p_0\varepsilon_k -{M'}^2  =  2 \vec{p}\;' \cdot \vec{k}= 2\sigma  p'kz  \]
with $\sigma=\pm1$ is the sign function defined in (\ref{signe}). A singularity in the $z$-variable would  exist at  $z=z_0$ given by
\begin{equation}\label{z_0}
  z_0(k) = \sigma\;  \frac{2p_0\varepsilon_k-{M'}^2}{2p'k}   
\end{equation}
provided $\mid z_0\mid \leq 1 $, that is for $k$ in the interval $  k_- \leq k \leq k_+$ with
\begin{equation}
k_{\mp}  =  \frac{1}{2} \left| p'  \mp p_0  \sqrt{1 - \left( \frac{2m}{{M'}} \right)^2  } \right|   \label{kmp}
\end{equation}
Notice, in particular, that this singularity exists only in the inelastic case since $2m<{M'}$. It is a moving singularity, depending on the value of the second argument $k$, as
well as on the momentum transfer $Q^2$ and the parameter $Q^2_c$.

\bigskip
To properly  account for this singularity  we split the $k$-integration interval  in three domains
\[  [0,+\infty]=  [0,k_-]  \cup [k_-,k_+]  \cup [k_+,+\infty] \]
as well as the corresponding integral (\ref{F2_2d})
\begin{eqnarray*}
F_{2}(Q^2) &=& I_1+I_2+I_3  
\end{eqnarray*}
with
\begin{eqnarray*}
ÃI_1 &=& \frac{i}{(2\pi)^3}\;  \int_0^{k_-}             k^2dk \; \int_{-1}^{+1} dz  \;  \bar{f}_{2} (z,k)  \cr 
 I_2&=& \frac{i}{(2\pi)^3}\;  \int_{k_-}^{k_+} k^2 dk \int_{-1}^{+1} dz \bar{f}_{2} (z,k)  \cr
ÃI_3&=& \frac{i}{(2\pi)^3}\;  \int_{k_+}^{+\infty}  k^2dk \; \int_{-1}^{+1} dz \;  \bar{f}_{2} (z,k)  
\end{eqnarray*}
\begin{itemize}
\item The integrals over $[0,k_-]$ and  $[k_+,+\infty]$ have a smooth integrand in both variables and the contributions $I_1$ and $I_3$ can be computed by standard methods.
\item The integral over $[k_-,k_+]$ has a singularity on at $z=z_0$. The integrand is regularized by using the usual subtraction procedure:
\begin{equation}\label{subs_g1}
  g_1(z,k)=  \left[ g_1(z,k)- \frac{g'_1(k)}{z-z_0}\right]     + 
 \frac{g'_1(k)}{z-z_0}  
 \end{equation}
with
\[ g'_1(k)=  \lim_{z\to z_0} (z-z_0)g_1(z,k) = {\rm Res}\;\left[ g_1(z,k)\right]_{z=z_0}   \]

To compute this quantity we take z from
\[   \epsilon^2_{p'-k}= m^2 +k^2  + {p'}^2- 2\sigma p'kz     =  \epsilon^2_{k}  + p_0^2-{M'}^2 -  2\sigma p'kz \quad\Longleftrightarrow\quad z= \sigma \frac{  \epsilon^2_k  + p_0^2 -{M'}^2 - \epsilon^2_{p'-k}} {2p'k}   \]
and together with (\ref{z_0}) we have
\[ z-z_0= \sigma \frac{ \epsilon^2_k  + p_0^2  - \epsilon^2_{p'-k}   -   2p_0\varepsilon_k}{2p'k}   = \sigma \frac{  (p_0-\varepsilon_k  )^2  - \epsilon^2_{p'-k} }{2p'k} \]
and so
\[ (z-z_0)g_1(z,k)  =  \sigma  \frac{G(+\varepsilon_k,z,k)}               { 4p'k\varepsilon_k  [ (p_0-\varepsilon_k )^2 - \epsilon^2_{p-k} ]   } \]
We get in this way the residue
\begin{equation}\label{gp_1}
   g'_1(k)= \frac{\sigma}{4p' k\varepsilon_k}  \;\left\{  \frac{  G(\varepsilon_k,z,k)}{ (p_0-\varepsilon_k+\varepsilon_{p-k}) \;(p_0 -\varepsilon_k-\varepsilon_{p-k})} \right\}_{z=z_0}   
 \end{equation}
\end{itemize}

\bigskip
The $g_6$ term  
\[    g_6(z,k) =     \frac{G(p_0-\varepsilon_{p'-k} ,z,k)} { 2\varepsilon_{p'-k}   (p_0-\varepsilon_{p'-k}+\varepsilon_k) ( p_0-\varepsilon_{p'-k} -\varepsilon_k)         \; (\varepsilon_{p-k} +  \varepsilon_{p'-k} ) (\varepsilon_{p-k} -  \varepsilon_{p'-k} )   }          \]
 has the same singularity  at $z=z_0$  than $g_2$ and  has been  treated in the same way by subtraction
 \[  g_6(z,k)=  \left[ g_6(z,k)- \frac{g'_6(k)}{z-z_0}\right]     + \frac{g'_6(k)}{z-z_0}  \]
 By a similar calculation we find 
\[  g'_6(k) =- g'_1(k)  \]

\bigskip
Finally, once regularized the singular terms $g_2$ and $g_6$,   the $I_2$ contribution is given by two integrals, corresponding to the two terms in the subtraction (\ref{subs_g1})
\begin{eqnarray}
I_2      &=&  I'_2 + I''_2  \\
I'_2 &=&  \frac{i}{(2\pi)^3}\;  \int_{k_1}^{k_2}      k^2dk \;\int_{-1}^{+1}  dz  \;  \bar{f}'_{2} (z,k)  \\
I''_4 &=&  \frac{i}{(2\pi)^3}\;  \int_{k_1}^{k_2}      k^2dk     \; c'_1(k) g'_1(k) \log \left|\frac{1-z_0}{1+z_0 }\right|    
\end{eqnarray}
where
\begin{eqnarray}\label{fbp_2} 
\bar{f}'_2 &=&  \bar{f}_2 -  c'_1(k)\frac{g'_1(k)}{z-z_0} 
\end{eqnarray}
is a regular  integrand, $g'_1$ the residue (\ref{gp_1}) and $c'_1(k)\equiv c_1(k)-c_6(k)$ with $c_i$  given in (\ref{c_i})

\subsection{$F_1$ contribution}\label{F_1}

The $F_1$ contribution is given by
\begin{equation}\label{F1b}
F_1(Q^2)=\frac{i}{(2\pi)^3} \; \int dk_0\; dz\; k^2dk \,f_1(k_0,z,k)  
\end{equation}  
According to definition (\ref{f1}) of $f_1$,  this function contains three contributions, each of them contains the product of two delta-functions.
It turns out that the non-zero contribution results from the second term only.

 We substitute (\ref{f1}) in (\ref{F1b}), integrate this term
over $k_0$ analytically  and find:
\begin{equation}\label{Fp1}
\tilde{f}_1(z,k)= \int dk_0 \,f_1 = -\frac{\pi^2}{4 \varepsilon_{\vec{k}}\varepsilon_{\vec{p'}-\vec{k}}}  \; \frac{G(\varepsilon_{\vec{k}},z,k)}{(\varepsilon_{\vec{p'}-\vec{k}}^2-\varepsilon_{\vec{p}-\vec{k}}^2)}   \; \delta(p_0-\varepsilon_{\vec{k}}-\varepsilon_{\vec{p'}-\vec{k}}) \;  
\end{equation}
to be integrated then over $z$ and $k$:
\begin{equation}\label{F1a}
F_1(Q^2) =\frac{i}{(2\pi)^3}   \int dz \;k^2dk \; \tilde{f}_1(z,k) 
\end{equation}

In general, the denominator $(\varepsilon_{\vec{p'}-\vec{k}}^2-\varepsilon_{\vec{p}-\vec{k}}^2)$ in (\ref{Fp1}) can vanish.
However, one can check that  if the argument of the delta function $(p_0-\varepsilon_{\vec{k}}-\varepsilon_{\vec{p'}-\vec{k}})$ is zero  --   providing a non-zero contribution --  
the denominator $(\varepsilon_{\vec{p'}-\vec{k}}^2-\varepsilon_{\vec{p}-\vec{k}}^2)$ is not zero.
Therefore there is no singularity from this denominator.

Concerning the first and the third terms in (\ref{f1}),  the arguments of the $\delta$-functions in them vs. $k_0$ can also cross zero.
However, in the first term they cannot be zero simultaneously. 
In the third term, the arguments of the $\delta$-functions can be zero simultaneously.  That is, after integration over $k_0$, we obtain the delta-function $\sim \delta(E_{\vec{p}-\vec{k}}^2-E_{\vec{p'}-\vec{k}}^2)$ which could contribute. However, a more careful analysis shows that its contribution is in fact zero. For this aim we represent the delta-function as
\[ \delta(x)=\frac{\epsilon}{\pi(x^2+\epsilon^2)} \]
and integrate  over both $dz$ and $dk$. Taking after that the limit $\epsilon\to 0$, we find  a  zero result. 
Care must be taken however to do not take this limit too early, i.e.,  before integration, since we 
will get in this way a wrong non-zero contribution. 
\bigskip

After integrating analytically over $dz$ in (\ref{F1a})  by means of the delta-function we obtain:
\begin{equation}\label{f1bar}
\bar{f}_1(k)=\int_{-1}^1 \tilde{f}_1 dz=\pi^2\theta(1-|z_0|)\left.\frac{1}{2 p'k \varepsilon_{\vec{k}}}\frac{G(k_0=\varepsilon_{\vec{k}},k)}{(\varepsilon_{\vec{p}-\vec{k}}^2-\varepsilon_{\vec{p'}-\vec{k}}^2)}\right|_{z=z_0}
\end{equation}
where the value of $z_0$ is given by (\ref{z_0}). 

\bigskip
Finally, the integration over $k$  is reduced, due to  the theta-function  \mbox{$\theta(1-|z_0|)$}, to the interval  $k\in[k_-,k_+]$ with $k_\mp$  given  (\ref{kmp}). That is:
\begin{equation}\label{F1fsi}
F_{1}(Q^2)=\frac{i}{(2\pi)^3}\int_{k_-}^{k_+}k^2dk\, \bar{f}_1(k)
\end{equation}
As a test, we carry out an independent  calculation for $\Gamma_i=\Gamma_f=1$,  using Feynman parametrization for the 4D integrals. In this way, we find the imaginary part which coincides with the contribution  (\ref{F1fsi}).

\subsection{About the function $G(k_0,z,k)$}\label{appB4}

An important quantity used in our formalism, which contains all the information about the initial and final state vertex amplitudes, is the function $G(k_0,z,k)$, defined in Eq. (\ref{Gk0}) 
\[ G(k_0,z,k)=\frac{(p_0-k_0)}{p_0} \; \Gamma_i \left(\frac{p}{2} -k,p\right)\Gamma_f \left(\frac{p'}{2}-k,p'\right)   \]
Some useful relation concerning this function are specified in what follows. 

\bigskip
The initial state amplitude  $\Gamma_i(k_0,k)$ -- we  denote $|\vec{k}|=k$   -- is computed in the reference frame where $\vec{p}=0$. 
In an arbitrary frame $\Gamma_i(k_0,k)$ is written as $\Gamma_i(\tilde{k}_0,\tilde{k})$
where
 \begin{eqnarray*}
 \tilde{k}_0  &=& \frac{k\cd p}{M } \cr
\tilde{k}       &=& \sqrt{  \frac{(k\cd p)^2}{M^2}-  k^2}
\end{eqnarray*} 
The arguments of  $\Gamma_i \left(\frac{p}{2} -k,p\right)$ are obtained from these relations by the shift 
$k\to \frac{p}{2}-k$. 
The same happens for the final state vertex amplitude $\Gamma_f \left(\frac{p'}{2}-k,p'\right)$. 
The function  $G(k_0,z,k)$ should be therefore  understood as:
\begin{equation}\label{G0}
G(k_0,z,k)=\frac{(p_0-k_0)}{p_0} \Gamma_i(\tilde{k}_0,\tilde{k})\Gamma_f(\tilde{k'}_0,\tilde{k'})
\end{equation}
with, after performing the shift $k\to \frac{p}{2}-k$, the arguments are given by
\begin{eqnarray*}
\tilde{k}_0 &=& \frac{M}{2}-\frac{k\cd p}{M}  \cr
\tilde{k}     &=& \sqrt{\frac{(k\cd p)^2}{M^2}-k^2}
\end{eqnarray*}
Written in  more detail:
\begin{eqnarray}\label{arg}
\tilde{k}_0  &=& \frac{M}{2} -\frac{k_0p_0-kpz}{M} \cr
\tilde{k}      &=& \sqrt{   \frac{(k_0 p_0-kpz)^2}{M^2}  -k_0^2+\vec{k}^2   }    \cr
\tilde{k'}_0  &=& \frac{M'}{2}  -  \frac{ k_0p_0-\vec{k}\cd\vec{p'} }{M'} \cr
\tilde{k'}      &=& \sqrt{\frac{(k_0 p_0-\vec{k}\cd\vec{p'})^2}{{M'}^2}-k_0^2+\vec{k}^2}
\end{eqnarray}
We remind that the sign in the scalar product $\vec{k}\cd\vec{p'}$ is given by  (\ref{k_dot_pp})  and  the values of $p,p'$ and $p_0$ are defined by Eqs. (\ref{ppp}) and (\ref{p0}). 
They all depend on the value of $Q^2$.

\bigskip
The initial (bound state) solution $\Gamma_i$ is normalized so that the elastic EM form factor at $Q^2=0$ is 1.
There is no any uncertainty in the normalization of the scattering state solution $\Gamma_f$ determined by the inhomogeneous BS equation. 
One should also take into account that the solution  found in \cite{CK_PLB_PRD} was the partial wave amplitude $F_0$ related to the full amplitude by
\[ \Gamma_f=16\pi\sum_{l=0}^{\infty}F_lP_l(\cos\theta)  \]
and that for the S-wave, the function $\Gamma_f$ in (\ref{G0}) is related to our solution $F_0$  obtained in  \cite{CK_PLB_PRD} by
\begin{equation}\label{GF0}
\Gamma_f=16\pi F_0.
\end{equation}

\section{Calculating the PW contribution to the transition form factor}\label{app_PW}

We carry out  here the part of integration over $d\Omega_{\vec{p}_s}$  and $d^4k$ in the form factor (\ref{Fpw})  that can be done analytically.
Integrating first over $d\Omega_{\vec{p}_s}$ in the frame $\vec{p'}=0$, $p'_0=M'=2\sqrt{m^2+\vec{p}_s^2}$ we get:
\begin{eqnarray*}
\int\delta^{(4)}\left(k-p_s-\frac{p'}{2}\right)\frac{d\Omega_{\vec{p}_s}}{4\pi}
&=& \int\delta\left(k_0-\frac{M'}{2}\right)\frac{1}{p_s^2}\delta(|\vec{k}|-p_s)
\delta^{(2)}(\Omega_{\vec{k}}-\Omega_{\vec{p}_s})\frac{d\Omega_{\vec{p}_s}}{4\pi}
\\
&=&\frac{1}{4\pi p_s^2}\delta\left(k_0-\frac{M'}{2}\right)\delta(|\vec{k}|-|\vec{p_s}|)
 \end{eqnarray*}
In an arbitrary frame $k_0$ and $|\vec{k}|$ are rewritten as
$$
k_0\to \frac{k\cd p'}{M'}, \quad |\vec{k}|=\sqrt{k_0^2-k^2}\to \sqrt{\frac{(k\cd p')^2}{{M'}^2}-k^2}
$$
Explicitly,  in the frame where $p'_0=p_0$:
\begin{eqnarray*}
\frac{k\cd p'}{M'}                               &=&  \frac{ k_0p_0 -  \sigma z|\vec{k}|p' }{M'}   \cr 
\sqrt{  \frac{(k\cd p')^2}{{M'}^2}  -k^2}  &=&  \sqrt{  \frac{(k_0p_0 - \sigma z|\vec{k}|p')^2}{{M'}^2}   -k_0^2+|\vec{k}|^2   }
\end{eqnarray*}
where the sign $\sigma$ is defined in (\ref{signe}).

After these transformations, the form factor Eq. (\ref{Fpw}) obtains the form:
\begin{eqnarray}\label{pw3}
F_{pw}&=&-\int  \frac{(p_0-k_0)}{p_0}\;\frac{\Gamma_i\left(\frac{p}{2}-k,p\right)}{[(p-k)^2-m^2+i\epsilon]}
\nonumber\\
&\times&
\delta\left(\frac{(k_0p_0- \sigma zkp'}{M'}-\frac{M'}{2}\right)
\delta\left(\sqrt{\frac{(k_0p_0 - \sigma  k p')^2}{{M'}^2}-k_0^2+k^2}-p_s\right)\,\frac{dk_0 k^2dk dz}{2p_s^2}
\end{eqnarray}
Eq. (\ref{pw3}) contains two-delta functions and integration over three variables  $k_0$, $k$ and $z$.
By means of the delta-functions  we integrate over the variables $k_0$ and $z$. If  $arg_1$ and $arg_2$ denote the arguments of the first and second delta-functions,
the conditions that  the arguments of both delta's equal to zero give the system of  equations $arg_1=0,\;arg_2=0$ which we solve relative to 
$k_0$ and $z$. We find that  $k_0=\varepsilon_k=\sqrt{m^2+k^2}$ and $z=z_0$ with the value of $z_0$ given by (\ref{z_0}).
Remind that $\mid z_0\mid \leq 1$  if $k\in [k_-,k_+]$ with $k_{\mp}$ given by (\ref{kmp}).

Calculating an integral containing the delta-function, one should divide the result by derivative, over the integration variable, of the argument of the delta-function. Since the integral (\ref{pw3}) contains two delta-functions,  we have to divide the result by the product  $d_1\,d_2$ of derivatives $d_1=arg'_1$, $d_2=arg'_2$ of the arguments of both delta-functions. Each derivative depends on the order of calculation (first over $k_0$, then over $z$ or in the opposite order), though the final results, which is determined by their product, is the same. That is:
$$d_1\,d_2=arg'_{1,k_0}\,arg'_{2,z}=arg'_{1,z}\,arg'_{2,k_0}=\frac{k \varepsilon_kp'}{M'p_s},$$ 
provided $arg_1=arg_2=0$.

Finally, the integral over $z$ is reduced to:
$$
\int_{-1}^1\ldots \delta(z-z_0)dz dk \sim \int\ldots \theta(1-|z_0|)dk=\int_{k_-}^{k_+}\ldots dk
$$
Thus,  after integration over $k_0$ and $z$ we obtain the integral (\ref{pw4}) over $k$ in the limits determined by the condition $|z_0|\leq 1$ and given by Eq. (\ref{kmp}).


\end{document}